# CrowdSource: Automated Inference of High Level Malware Functionality from Low-Level Symbols Using a Crowd Trained Machine Learning Model

Joshua Saxe
Invincea Labs
Josh.saxe@invincea.com

Rafael Turner
Invincea Labs
Rafael.turner@invincea.com

Kristina Blokhin
Invincea Labs
Kristina.blokhin@invincea.com

## ABSTRACT

In this paper we introduce CrowdSource, a statistical natural language processing system designed to make rapid inferences about malware functionality based on printable character strings extracted from malware binaries. CrowdSource "learns" a mapping between low-level language and high-level software functionality by leveraging millions of web technical documents from StackExchange, a popular network of technical question and answer sites, using this mapping to infer malware capabilities. This paper describes our approach and provides an evaluation of its accuracy and performance, demonstrating that it can detect at least 14 high-level malware capabilities in unpacked malware binaries with an average per-capability f-score of 0.86 and at a rate of tens of thousands of binaries per day on commodity hardware.

## 1. INTRODUCTION

To understand the cyber threat landscape the thousands of new malware variants observed every month must be inspected. Thus it would be useful to use automation to accelerate the reverse engineering process to scale human analysts' efforts to large volumes of malware.

While existing literature on automated malware analysis has proposed methods for identifying similarity relationships between malware artifacts [1]–[4], and methods for automatically classifying malware into known malware families [4], [5], far fewer methods have been proposed for rapidly generating "capability profiles" for malware binaries [6]. A key reason for this lack of work on capability detection for malware, we believe, is that abstraction from low-level malware features has mostly been thought of as a manual knowledge engineering problem.

Indeed, the approach taken by currently deployed automatic malware analysis systems, to our knowledge, is almost entirely based on hand-coded rules defined on statically and dynamically obtained features such as API calls and protocol strings [7]. While such approaches can be powerful in the hands of experts, a limitation of these approaches is that while they perform automatic inference on malware, the knowledge engineering work they require is anything but automatic and is hard to scale to the thousands of possible technical symbols that predict malware functionality of interest.

In this paper we propose a new and complementary approach to existing automated malware analysis techniques, presenting a statistical capability detection model learned from millions of programming-related documents drawn from the StackExchange network of technical question and answer websites. Figure 1 gives intuition for the CrowdSource approach, showing how a post on the site StackOverflow.com containing the capability relevant keyword "screenshot" also contains terms (such as "FindWindow" and "PrintWindow") that occur in malware and are indicative of "screenshot" functionality. Our intuition is that the co-occurrence of high level language like "screenshot" with low-level API calls used to implement this functionality can be exploited to detect capabilities such as "screenshot grabbing" within malware. To do this, we have developed a Bayesian network based statistical approach for mapping large vocabularies of terms extracted from StackExchange to the high level capabilities they indicate.

Because we have evaluated our approach, called CrowdSource, on malware printable strings data, we note that our approach requires deobfuscated (or "unpacked") representations of malware binaries if it is to succeed in achieving useful inferences. For the purposes of this paper, we do not engage the related and important problem of malware deobfuscation.

**Informal overview of the CrowdSource approach**

The goal of CrowdSource is to detect high-level malware capabilities based on malware strings data. Here we define capabilities to be high-level functionality of interest to a malware analyst or network defender. Example capabilities are "implements keystroke logging," "turns on webcam," and "takes screenshots of user's desktop."

We have chosen to use printable strings extracted from malware to detect malware capabilities because, when malware is not obfuscated or is de-obfuscated, many

Work funded under DARPA contract FA8750-10-C-0169. Approved for public release, distribution unlimited. The views expressed are those of the authors and do not reflect the official policy or position of the Department of Defense or the U.S. Government.

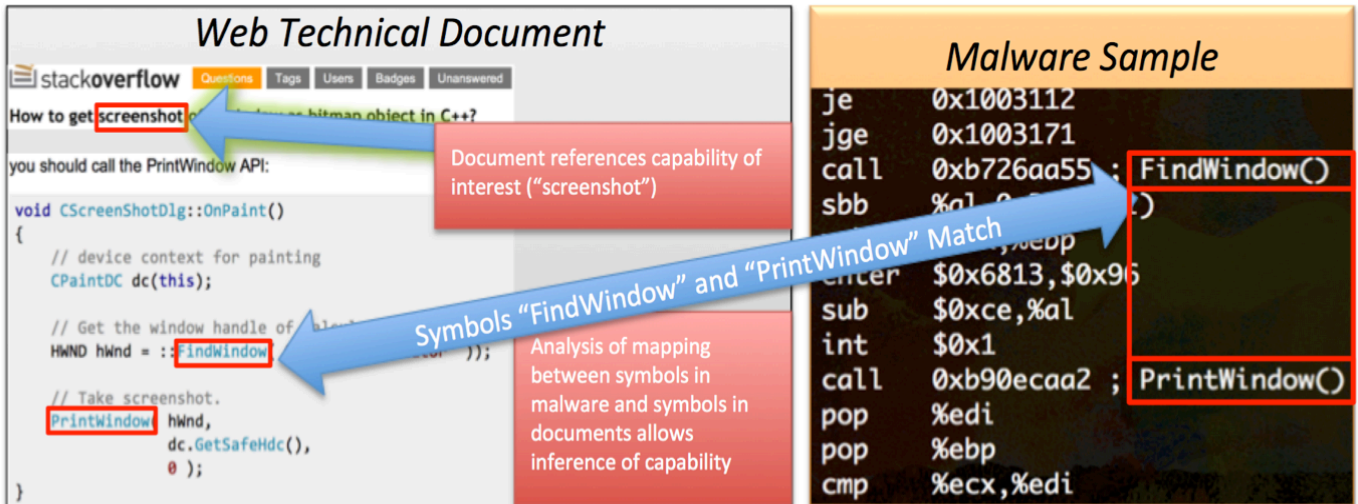

**Figure 1.** An illustration of the intuition that we can correlate terms found in malware binaries and terms found in StackExchange documents to identify high-level software capabilities within malware

malware variants' strings data contain the API symbols malware calls, the DLL files malware imports, registry keys malware touches, and protocol format strings malware uses to communicate with other hosts, and thus these data form a rich starting place for applying natural language processing methods to malware capability inference. While we evaluate CrowdSource on strings data here, we believe our approach could also have applicability to other malware data, such as dynamic trace data.

Figure 2 gives an overview of the CrowdSource approach to detecting capabilities in malware binaries. As shown in Figure 2, to use CrowdSource to detect capabilities within malware, the user creates a capability configuration file defining what capabilities they are interested in, also creating a search query for each capability, thereby telling CrowdSource how to retrieve StackExchange documents relevant to that capability. For example, to use CrowdSource to detect malware that takes screenshots, the user might use the query "screenshot implementation," leading CrowdSource to retrieve documents about implementing screenshot functionality.

As further illustrated by Figure 2, CrowdSource uses the documents that are returned by each capability search query to find terms that are probabilistic indicators for the capability. When CrowdSource observes these terms in malware samples strings data the system is able to compute a probability that the malware sample has a given capability.

By enabling this workflow for any number of capabilities, CrowdSource provides capability "profiles" for malware binaries. Furthermore, because our capability inference is a very fast process, taking about 200 milliseconds per sample on average, CrowdSource can perform this analysis on hundreds of binaries per minute depending on the user's hardware configuration.

Below we describe CrowdSource's system components, and then give a detailed description of the workflow by which we initialize CrowdSource and detect capabilities within malware.

## 2. SYSTEM OVERVIEW

In this section we describe CrowdSource's four components and their interoperation. First we give succinct definitions of each component:

**1) The full-text document index:** the component we use to store the programming question and answer documents from StackExchange.

**2) The capability configuration:** A configuration file where the user defines what capabilities he or she is interested in identifying within malware and gives search queries for these capabilities. These search queries are used to identify documents within which we identify terms that are probabilistically indicative of capabilities.

**3) The term-capability conditional probability matrix**: A matrix which holds automatically identified probabilistic associations between terms and capabilities that are used to compute the probabilities that a malware sample has a given set of capabilities.

**4) The malware capability inference module:** a module that extracts terms from malware strings data and leverages the *term-capability conditional probability matrix* to infer the probability that a malware binary has each of the capabilities defined in the *capability configuration*.

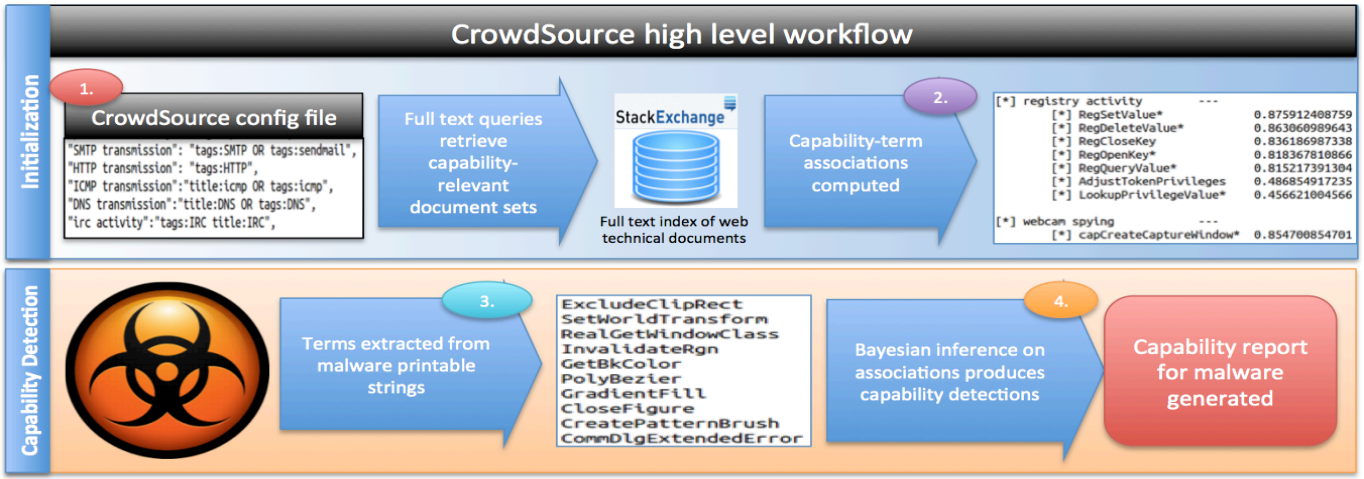

**Figure 2.** An overview of the CrowdSource workflow.

Below we describe these components and their interoperation in detail.

### The full text document index

We use the *full text document index* to index a corpus of StackExchange programming question and answer documents that we use as a knowledge base for CrowdSource. Each question in the StackExchange corpus constitutes a discussion thread, where a question includes a *question title* (typically a sentence length version of the technical question), a variable number of *question tags* (user defined tags that associate the question with technical topics like "opengl") a *question document* (the question written in longer form) and some variable number of *answer documents*.

We index these documents in the *full text document index*. All documents on the StackExchange sites, including both the single question document that heads each question thread, and each of the variable number of answer documents attached to each thread, are indexed as individual documents with a title, tags and body field. We store all of this text in *markdown* format [8] such that HTML metadata is stripped and all that remains is the literal text of each field.

### The capability configuration

The *capability configuration* is the way that the CrowdSource user specifies the capabilities they want CrowdSource to detect and seeds CrowdSource with search queries that retrieve documents relevant to these capabilities. CrowdSource then uses these documents to find probabilistic indicators of these capabilities.

These search queries are defined as Boolean conditions on the documents' terms. Thus to find documents related to "remote desktop protocol" access to a host, we might define a query such as "title:rdp AND tags:rdp," which matches documents that have the term "rdp" in both their title and tags.

### The term-capability conditional probability matrix

CrowdSource computes the probability that malware binaries have capabilities based on probabilistic indicator terms found in the malware strings. The *term-capability conditional probability matrix* stores these indicators and their conditional probabilities, which give the probability that a capability is present if the indicator term is present.

To make the concept of a probabilistic indicator term concrete consider the case of the "communicates via Internet Relay Chat" (IRC) capability. The term "PRIVMSG" is highly indicative of IRC capability because "PRIVMSG," an IRC protocol directive, almost always occurs in documents about IRC and almost never occurs elsewhere. CrowdSource thus assumes that when malware has the term PRIVMSG, it has a high probability of *implementing* IRC communications.

The *term-capability conditional probability matrix* stores such probability values for every term in the capability search query result sets. The matrix contains three columns: a term column, a conditional probability column, and a capability name column. For each term $T_j$ found in the documents matching a search query for a capability $C_i$, the conditional probability that the capability is present given the term is computed by the following formula:

(1) $$\forall T_j \in C_i;\ P(C_i|T_j) = \frac{|T_{j_c}| + \alpha}{|T_{j_t}| + \alpha + \beta}$$

Here $T_{j_c}$ is the set of occurrences of a given term within posts describing the target capability, $T_{j_t}$ is the set of occurrences of the term in the *full-text document index* overall, and $\alpha$ and $\beta$ are hyperparameters that allow us to adjust our prior belief that any given term in a result set should either indicate a given capability or not indicate the capability.

| Dataset | Questions | Answers |
|---|---|---|
| StackOverflow | 5.4M | 9.9M |
| SuperUser | 181K | 307K |
| ServerFault | 156K | 291K |
| AskUbuntu | 111K | 143K |

**Fig. 3.** Document training datasets used in our evaluation.

We note that (1) constitutes a Beta-Bernoulli model for estimating the Bernoulli parameter [9]. $\alpha$ and $\beta$ are the parameters of the Beta, and can be modified to incorporate prior beliefs about the degree to which a given word indicates a given capability. For our purposes we use the $\alpha$ and $\beta$ hyperparameters to model our prior belief that words observed at least once in the query documents indicate the capability the query documents describe.

For example, for any word in the query matching documents, we can set $\alpha$ and $\beta$ to 10 and 90 respectively, giving a 10% prior probability for $P(C_i|T_j)$ for words that have been observed at least once in the context of a discussion of the capability, which will be balanced with the empirical observations modeled by $T_{j_c}$ and $T_{j_t}$. For the purposes of this paper, we set $\alpha$ and $\beta$ globally, to 10 and 90 respectively as we have found these parameters give us the best results.

While understanding why exactly this is the case will require additional investigation, our initial hypothesis is that it is because it makes up for both precision and recall bias in our document retrieval approach. If it is true that our document retrieval approach generally retrieves some documents that are *not* about the capability while also missing some documents that *are* about the capability, then global priors such as those we have used here may possibly help adjust for this bias.

*The malware capability inference module*

The goal of the *malware capability inference module* is to compute the final probability that malware binaries have capabilities of interest. As mentioned above, to compute these final probabilities we use a Bayesian network approach for combining independent indicator term conditional probabilities into a final probability. Bayesian networks allow us to reason in a principled manner about the following question: given a set of probabilistic indicator terms extracted from a malware binary (say "privmsg", "topic", and "channel" for IRC), what is the final probability that the malware binary does indeed have this capability?

In detail, this approach works as follows. To extract terms from malware, we identify all printable substrings of the binary file greater than length four and then tokenize these strings along non-alphanumeric boundaries. The resulting tokens are the malware's *terms*. Having extracted a set of terms from the malware binary (which can be natural language words, API call names, etc.), we then lowercase normalize the terms. For each capability $C_i$ in the *capability configuration* and for each of the malware terms $T$ it is associated with, we compute a final probability the capability is present $P(C_i)$, as follows:

(2) $$P(C_i|T_j \ldots T_n) = 1 - \prod_j (1 - P(C_i|T_j))$$

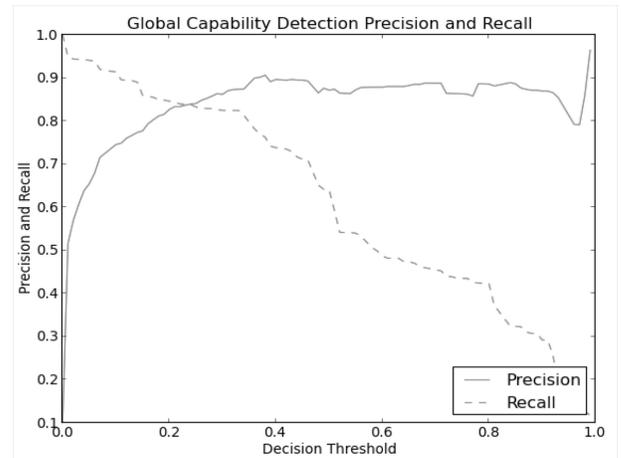

**Fig. 4.** Precision and recall in relation to threshold

In the Bayesian network formalism, (2) defines a "noisy-OR" gate between the observed words in the malware and the probability that the malware contains a given capability. Noisy-OR gates such as (2) model the Boolean "OR" operation in a probabilistic context, asking the question, what is the probability that at least one of some enumeration of independent, probabilistically defined Boolean conditions is in fact true [10].

To make this idea concrete, consider a hypothetical case in which a malware binary has the terms "privmsg", "topic", and "channel" and we seek to compute the probability that the binary has the capability "IRC." Assuming P(IRC|privmsg) = 0.9, P(IRC|topic) = 0.5, and P(IRC|channel)=0.1, (2) would evaluate to 0.955. In other words, (2) dictates that a malware capability is at least as probable to be present as the most indicative term, and grows in probability with each additional probabilistic indicator term.

Once the *malware capability inference module* has computed probabilities for each capability in the *capability configuration* CrowdSource can optionally discretize these probabilities into binary capability detections ("yes" or "no" determinations that the binary does or does not have given capabilities) using a decision threshold. Alternatively, we can inform the user of the computed probabilities. For the purposes of evaluation we make a binary decision to produce binary capability detections, as we detail in the next section.

## 3. EVALUATION

In the experiments described in this paper we initialized CrowdSource on a document corpus acquired from the StackExchange network of technical question and answer websites. The StackExchange organization releases its question and answer documents under the Creative Commons license and in an easily parseable XML format. The table in Figure 3 describes the contents of each document corpus, illustrating that we trained our model on over 16 million documents.

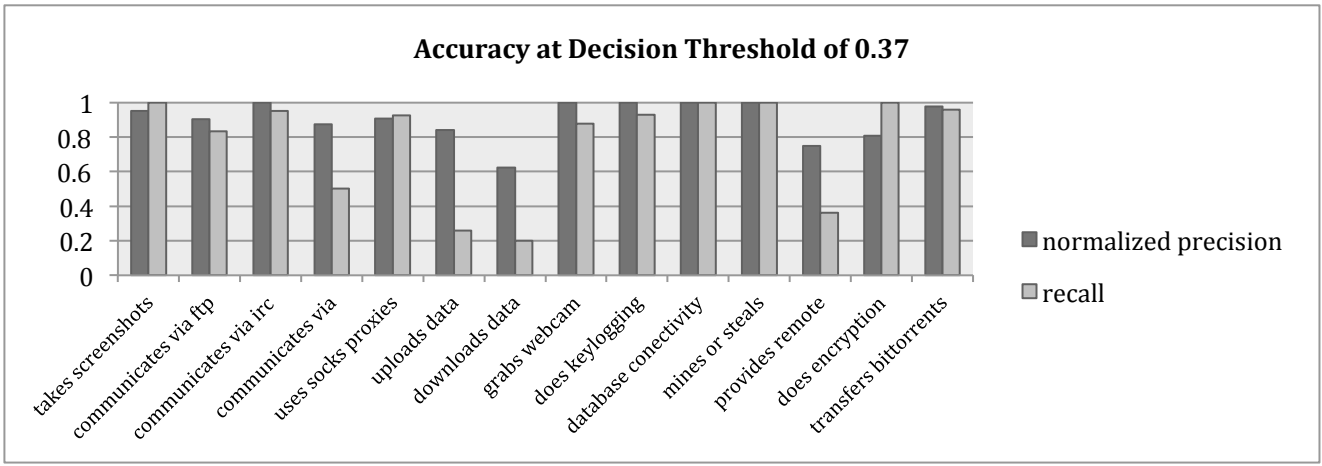

**Fig. 5.** CrowdSource accuracy against a test corpus of labeled malware at threshold 0.37

To evaluate CrowdSource, we hand-reverse engineered 1457 malware binaries and labeled each as implementing any of 14 capabilities. We obtained these malware binaries from two sources: first, from the MD:Pro malware feed from Frame4 [11], and second, from a team at MIT Lincoln Laboratories which provided the binaries to us for testing of the automated malware analysis prototypes we are developing in our lab.

For each of these malware binaries, we verified that the binary was not "packed" (transformed by a binary obfuscator). After verifying that the malware was unpacked, we manually inspected each binary to determine which capabilities it implements. We chose not to label any malware binaries as not implementing a given capability, because verifying that a program does not do a given behavior is an often intractable and error prone endeavor. Instead, to provide "negative" test examples for our evaluation, we assembled a set of 377 benign Windows XP binaries included with the Windows operating system. For each of our 14 capabilities, we created a "negative" test set composed of a subset of these 377 benign binaries that we verified, by inspecting the Windows XP documentation, do not implement the capability of interest.

After creating our ground truth evaluation dataset, we created a *capability configuration* containing these 14 capabilities and their corresponding search queries over our *full-text document index*. The process by which we created the search queries in this configuration was one of trial and error, by which we attempted to formulate queries that maximized the number of relevant documents that were returned by the queries and minimized the number of irrelevant documents returned by the queries. We found that finding good queries is a time-consuming process: the optimization of CrowdSource's *full-text document index* to allow users to quickly and easily identify the documents that describe the capability they are interested in would thus be important future work.

The metrics we use to evaluate CrowdSource's accuracy once we have decided on a *capability configuration* are a normalized variant of precision, and recall. These metrics are based on the concepts of false and true positives and false and true negatives. In our case, a false positive is a case where CrowdSource claims that a binary has a capability when in fact it does not, a false negative is a case where CrowdSource claims a binary does not have a capability where it really does, a true positive is where CrowdSource correctly detects a capability and a true negative is a case where CrowdSource did not detect a capability which it should not have.

Precision is defined as follows:

$$(2) \quad \frac{|True\ Positives|}{|True\ Positives|\ +\ |False\ Positives|}$$

Whereas recall is defined as follows:

$$(3) \quad \frac{|True\ Positives|}{|True\ Positives|\ +\ |False\ Negatives|}$$

As such, precision is sensitive to the distribution of positive and negative examples in one's test data. In our case, because we have varying distributions of positive and negative test examples for malware capabilities, we modify precision so that we can it to be invariant to the underlying ratio of positive to negative examples for each capability. We normalize it so that they each represent precision and recall over an invariant ratio between true and false test examples, using the following equations:

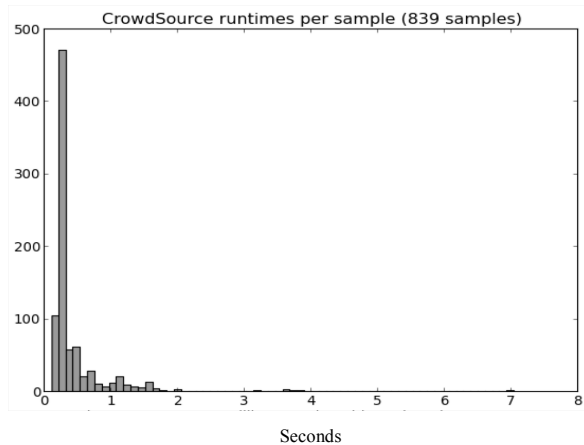

**Fig. 6.** Histogram giving runtimes for capability inference on 839 test binaries.

True positive rate (TPR):

$$(4) \quad \frac{|True\ Positives|}{|True\ Positives| + |False\ Negatives|}$$

False Positive Rate (FPR):

$$(5) \quad \frac{|False\ Positives|}{|False\ Positives| + |True\ Negatives|}$$

Normalized Precision:

$$(6) \quad \frac{TP * TPR}{TP * TPR + (1 - TP) * FPR}$$

For the purposes of this paper we set "true percentage" (TP) to 0.5, meaning that a given malware binary has a 50% chance of either having or not having the capability in question. Whether or not this is a realistic distribution will depend on the context in which CrowdSource is used; we chose this number because it is neutral with respect to the ratio between positive and negative binaries.

Figure 4 displays CrowdSource's normalized precision and recall as we vary a decision threshold from 0 to 1. Two potentially optimal points can be identified in the figure: the area around 0.25 and 0.37. At 0.25, the precision and recall lines cross at around 0.8, meaning that 20% of CrowdSource's capability detections are false positives, and 20% are false negatives, over all 14 capabilities which we evaluate here. At 0.37, CrowdSource's error is reduced in precision (false positives) and increased in recall (false negatives, or missed detections). Depending on use case, an operator may want to pick any number of decision thresholds. Figure 5 displays a breakout of CrowdSource's per-capability accuracy at a threshold of 0.37.

The CrowdSource model is efficient to initialize for new capabilities, and also performs fast inference of capabilities when run on malware binaries. To test the 14 capabilities evaluated here, CrowdSource required 119 seconds, on average (with a standard deviation of 91 seconds), to learn each capability, on a single core 2Ghz x86 processor with 32GB of RAM. In terms of computational complexity, this process is O(n), with n as the number of unique words in the documents matching the capability query.

Figure 6 gives empirical runtimes for CrowdSource when performing inference on malware, illustrating that CrowdSource performs efficient inference on malware and is a suitable tool for analyzing tens of thousands of binaries per day. As is clear from the figure, the most common runtime for malware binary capability inference is approximately 200 milliseconds, with an observed upper bound of 7 seconds and lower bound in the 100-millisecond bin. The system infers malware capabilities in linear time with respect to the number of terms extracted from each malware binary as can be seen in (2).

**Qualitative aspects of our approach**

A property of our system that is more difficult to measure empirically is the traceability it provides for its detection decisions, which we believe may help analysts reverse engineer malware more quickly and effectively. Indeed, by allowing analysts to see *why* CrowdSource has flagged a binary as having some likelihood of having some capability, we believe we can give them more insight into whether or not a binary has a capability, and, if the binary does have the capability, how it is implemented.

Figure 7 shows command line output from our research prototype. Our goal in designing the output of the command line tool was, for each symbol associated with each capability detected within a malware binary, to give evidence for why that symbol gives evidence that the capability detection is correct. In other words, we attempted to "auto-document" CrowdSource's output for the user, so as to give more experienced users justification for the tool's decisions and less experienced users both justification and background information which they may use to learn the meaning of technical symbols they haven't encountered before.

As shown in the Figure, we do this in two ways. First, we give an example post title that is "about" the capability of interest and also contains the symbol found in the malware. Second, we also give an example post title and we also give a text snippet from within that post to highlight where the symbol found in the malware occurs in the post.

## 4. RELATED WORK

A significant body of work within both academic research and the open source security software community overlaps with what we have proposed in this paper. The closest effort to CrowdSource is described in [6]. This work came out of our lab and preceded the effort described in the present paper. [6] was an early effort to use web technical

```
[*]  low level GUI calls / screenshot (somewhat likely)
  [*] Extracted from sample:                Post title on StackOverflow

  [-] 'capturing screen'                    Capturing screenshot in iphone?

  [|] s code but it not working for me .... it gave me same result. :( upsate - hello RRB:
  [|] this code is working into my application for capture the screenshot of a iphone. -
  [|] can i give the demo for that - actually i want to create video of game play using
  [|] capturing screen shots along with sound. I want functionality like talking
  [|] tom application. - you should probably take a look at AVFoundation and specifically
  [|] AVAssetWriter as a way of creating videos of your screen content. - Take the
  [|] screenshot of yo

  [-] 'BitBlt'                              Capturing screenshot

  [|] // bitmap handle HBITMAP hbitmap = CreateCompatibleBitmap(hdc_screen, bounds.width,
  [|] bounds.height); // select the bitmap handle SelectObject(hdc_memory, hbitmap); //
  [|] paint onto the bitmap BitBlt(hdc_memory, bounds.x, bounds.y, bounds.width,
  [|] bounds.height, hdc_screen, bounds.x, bounds.y, SRCPAINT); // release the screen DC
  [|] ReleaseDC(NULL, hdc_screen); // get the pixel data from the bitmap handle and put it

  [-] 'CreateCompatibleDC'                  Capturing screenshot

  [|] void get_screenshot(COLORREF** img, const Rectangle &bounds) { // get the screen DC
  [|] HDC hdc_screen = GetDC(NULL); // memory DC so we don't have to constantly poll the
  [|] screen DC HDC hdc_memory = CreateCompatibleDC(hdc_screen); // bitmap
  [|] handle HBITMAP hbitmap = CreateCompatibleBitmap(hdc_screen, bounds.width,
```

**Figure 7. Terminal output from the CrowdSource prototype displaying a capability detection on a "KBot" IRC bot sample in our test corpus. Here the tool has been set to a high "verbosity" level, showing which posts contained the symbol identified in the malware but also where in those postings the symbols occurred.**

documents to gain insight into malware symbols, and describes a method for using symbol co-occurrence patterns in web question and answer posts to group semantically related symbols extracted from a malware binary into clusters.

Additionally, [6] describes a method for associating StackOverflow folksonomy tags with the binaries so as to make unsupervised malware capability detections. What is different about the work described in this paper is that the approach described here is semi-supervised; instead of attempting to gain insight into malware binaries based on the tags they are associated with on StackOverflow, we allow users to define queries into a large corpus of question and answer site postings so as to define the capabilities they are looking for. The result is a more accurate system with a less noisy system output format that allows users to define in advance the capabilities they would like to detect within malware.

Another related effort is an open source project, RE-Google, a plugin for the popular commercial reverse engineering framework IDAPro [12]. RE-Google accelerates the reverse engineering process by mapping between textual strings found in malware and open source source code stored on Google Code. Similar to RE-Google is RE-Source, which maps between features in malware binary code and a large array of Internet source code search engine APIs [13]. The difference between CrowdSource and RE-Google and RE-Source is that while RE-Google and RE-Source identify code deemed to be similar to disassembled malware code, CrowdSource produces binary capability classifications, predicting whether or not discrete capabilities exist in the malware. And whereas RE-Google and RE-Source are based on textual source code, CrowdSource trains on question and answer documents.

The Yara framework [7], as discussed above, relates to CrowdSource in that it provides a simple domain specific language which can be used to identify malware capabilities. Similarly, [3] describes a system that uses expert rules in the capability identification module in order to classify malware binaries based on what traits they have where traits are abilities like accessing the webcam or taking screenshots. Of course, unlike Yara and the work described in Saxe et al., CrowdSource does not rely on expert rules to detect malware capabilities.

In the domain of machine learning, Yavvari et al. present a system that identifies the behavioral components of malware done by clustering extracted features from a dynamic execution run in a sandbox [1]. While identifying behavioral components is related to the problem of detecting malware capabilities, Yavvari's components are unlabeled and thus give no information as to the functionality they actually produce.

CrowdSource is based on applying natural language processing techniques to the software engineering domain and in this area it is not alone. Mokhov et al. applied NLP and machine learning (ML) techniques to static code analysis to discover software vulnerabilities in benign programs by comparing a database of vulnerabilities against the programs source code. NLP techniques have also been applied to source code analysis by Kuhn et al. [14], although their purpose was to cluster software that had similar purposes.

Applying natural language processing techniques to question and answer websites like StackOverflow is a recent trend in the research literature. The research presented by [15] explores the community aspect of these websites in order to determine how valuable answers are. This is important to CrowdSource as there is the possibility of data misleading the outcome of our capability definitions. The research draws the conclusion that not all questions may be sufficiently answered but also concludes that StackOverflow has little problem with user misbehavior due to their reputation system which allows users of the system to assign low reputation scores to postings of low utility.

Some research has attempted to draw on human analysis in order to gain insight into malware capabilities. A crowdsourcing toolkit was developed by [16] for the purpose of threat detection during web browsing. The toolkit is a system based on machine learning and natural language processing that incorporates user input in order to categorize websites based on their detected level of maliciousness. Unlike our system, which relies on a corpus of documents to learn malware capability detection models, their system depends on user participation to detect malicious websites.

CrowdSource implements Bayesian based probabilistic methods for NLP and there have been multiple security detection research efforts that employ similar approaches. The most relevant case is the work of Villamarin et al. [17]

where a method for botnet detection was introduced. They derive their model from previous works on email analysis relating to spam detection where certain words and their frequencies predict spam. Their work looks at DNS traffic as features in detecting botnet clusters using a Bayesian network similar to CrowdSource. We look at symbolic information to detect capabilities whereas [17] looks at DNS traffic to detect botnet activity.

## 5. CONCLUSION AND FUTURE WORK

In this paper we have proposed and evaluated a statistical capability detection model estimated with a corpus of millions of web technical question and answer documents. We have shown that we are able to automatically perform capability inference by matching symbols extracted from malware binaries to the symbols that constitute the model.

CrowdSource is currently actively used in our own lab and in use at two government agencies, giving evidence that the approach has real merit for its ability to help today's malware analysts automate a portion of their reverse engineering workflow. The tool is also available on our company website [18], licensed as freeware.

Given the early success of our natural language processing approach against malware strings data, we believe a fruitful line of investigation would involve applying similar techniques to malware dynamic trace data. Additionally, our approach might be extended to more in-depth static analysis approaches, including C++ symbol demangling and analyzing string references in the context of control flow analysis so as to infer where malware implements detected functionality, for example.

## 6. ACKNOWLEDGEMENTS

We would like to thank the anonymous reviewers for their very helpful comments. We would also like to thank Jose Nazario, Giacomo Bergamo, Aaron Liu, Robert Gove, David Slater, and Konstantin Berlin for their helpful feedback. Finally, we would like to thank the Defense Advanced Research Projects Agency (DARPA) for the financial support that made this work possible.